\documentclass{IEEEtran}
\usepackage{amsmath,amsfonts}
\usepackage{algorithmic}
\usepackage{algorithm}
\usepackage{array}
\usepackage[caption=false,font=normalsize,labelfont=sf,textfont=sf]{subfig}
\usepackage{textcomp}
\usepackage{stfloats}
\usepackage{url}
\usepackage{verbatim}
\usepackage{graphicx}
\usepackage{cite}
\begin{document}

\title{A Novel Signal Detection Method for Photon-Counting Communications with Nonlinear Distortion Effects}

\author{Chen Wang, Zhiyong Xu, Jingyuan Wang, Jianhua Li, Weifeng Mou, Huatao Zhu, Jiyong Zhao, Yang Su, Yimin Wang, and Ailin Qi
\thanks{This work was supported by National Natural Science Foundation of China (Grant No. 62271502, No. 62171463, and No. 61975238) and Natural Science Foundation of Jiangsu Province (Grant No. BK20231486). (\textit{Corresponding author: Jingyuan Wang}.) 

Chen Wang, Zhiyong Xu, Jingyuan Wang,  Jianhua Li, Jiyong Zhao, Yang Su, Yimin Wang, and Ailin Qi are with the College of Communications Engineering, Army Engineering University of PLA, Nanjing 210007, China (e-mail: 0910210239@njust.edu.cn; njxzy123@163.com; 13813975111@163.com; 18021528752@163.com; zhaojiyong@whu.edu.cn; qieziyangyang@163.com; vivhappyrom@163.com; qial1212@163.com).

Weifeng Mou, Huatao Zhu are with the College of Information and Communication, National University of Defense Technology, Wuhan 430010, China (e-mail: weifengmou@126.com; zhuhuatao2008@163.com). }
\thanks{Manuscript received ; revised .}}

\markboth{Journal of \LaTeX\ Class Files,~Vol.~xx, No.~xx, xx~xx}%
{Shell \MakeLowercase{\textit{et al.}}: A Sample Article Using IEEEtran.cls for IEEE Journals}

\IEEEpubid{}

\maketitle

\begin{abstract}
This paper proposes a method for estimating and detecting optical signals in practical photon-counting receivers. There are two important aspects of non-perfect photon-counting receivers, namely, (i) dead time which results in blocking loss, and (ii) non-photon-number-resolving, which leads to counting loss during the gate-ON interval. These factors introduce nonlinear distortion to the detected photon counts. The detected photon counts depend not only on the optical intensity but also on the signal waveform, and obey a Poisson binomial process. Using the discrete Fourier transform characteristic function (DFT-CF) method, we derive the probability mass function (PMF) of the detected photon counts. Furthermore, unlike conventional methods that assume an ideal rectangle wave, we propose a novel signal estimation and decision method applicable to arbitrary waveform. We demonstrate that the proposed method achieves superior error performance compared to conventional methods. The proposed algorithm has the potential to become an essential signal processing tool for photon-counting receivers.
\end{abstract}


\section{Introduction}
\IEEEPARstart{I}{n} recent years, there has been a growing interest in using single-photon detectors to enhance the receiver sensitivity in optical wireless communication (OWC) \cite{ref1,ref2,ref3}. In photon-starved applications, the received optical signal often falls below the sensitivity of traditional photodetectors, such as p-i-n diodes or avalanche photodiodes (APDs), leading to signal loss due to thermal noise. To achieve photon-sensitive reception, single-photon detectors and photon-counting technology are being implemented in OWC systems. Among the various single-photon detectors available, the single-photon avalanche diode (SPAD) has become popular in the near-infrared and visible light spectrum, because of its compact size, high stability, and all-solid structure \cite{ref4}. SPADs offer significant avalanche gain, which allows them to overcome gain-dependent excess noise and intrinsic thermal noise, thereby achieving single-photon sensitivity \cite{ref5}.

The performance of OWC systems that use photon-counting receivers is influenced by the non-perfect factors of SPAD, including dead time, non-photon-number-resolving, and afterpulsing effect \cite{ref6}. Dead time refers to the finite period, typically several nanoseconds, during which the detector is inactive and unable to detect photons. SPAD can only capture a fraction of the optical signal during the gate-ON interval, discarding incident photons during the dead time. Previous studies often assume that the received waveform is an ideal rectangle wave \cite{ref7,ref8}. However, the limited bandwidth of light source, and channel effects like atmospheric turbulence, can cause the impulse response to deviate from an ideal rectangular shape \cite{ref9,ref10}. Unlike conventional linear photodetectors, the non-photon-number-resolving property of SPAD makes them nonlinear detectors, meaning that different waveforms with the same optical intensity can result in different photon counts. Therefore, for practical photon-counting receivers, it is critical to accurately estimate the signal and make decisions, taking into account the pattern of waveform. 

\section{System Model and Problem Formulation}
\subsection{Arbitrary Waveform Detection}
In time-gated mode, the SPAD is regularly armed \cite{ref11} and can detect photons only during a specified gate-ON interval ${\tau _{\rm{g}}}$. SPAD is inactive during the dead time ${\tau _{\rm{d}}}$. Define the detection cycle as ${\tau _{{\rm{cyc}}}} \buildrel \Delta \over = {\tau _{\rm{g}}} + {\tau _{\rm{d}}}$. The detected photon counts is the number of avalanche events within a symbol duration ${T_{\text{s}}}$.

The symbol information is obtained by opening several gates within the symbol duration and comparing the detected photon counts with a threshold \cite{ref12}. Fig. 1 illustrates how a single SPAD opens multiple gates within the symbol duration, transforming the continuous optical pulse into discrete avalanche events. Due to the non-rectangle waveform, the trigger probabilities for different gates vary. Since the optical power is concentrated in the middle of the symbol duration, the central gate has a higher trigger probability than the edge gates. Additionally, for practical SPAD arrays, the photon detection efficiencies (PDEs) of pixels at different positions are non-uniform \cite{ref13,ref14}, leading to different trigger probabilities. Consequently, the detected photon counts for practical photon-counting receivers deviate from the binomial distribution previously assumed in earlier works. 

\begin{figure}[!t]
\centering
\includegraphics[width=3.0in]{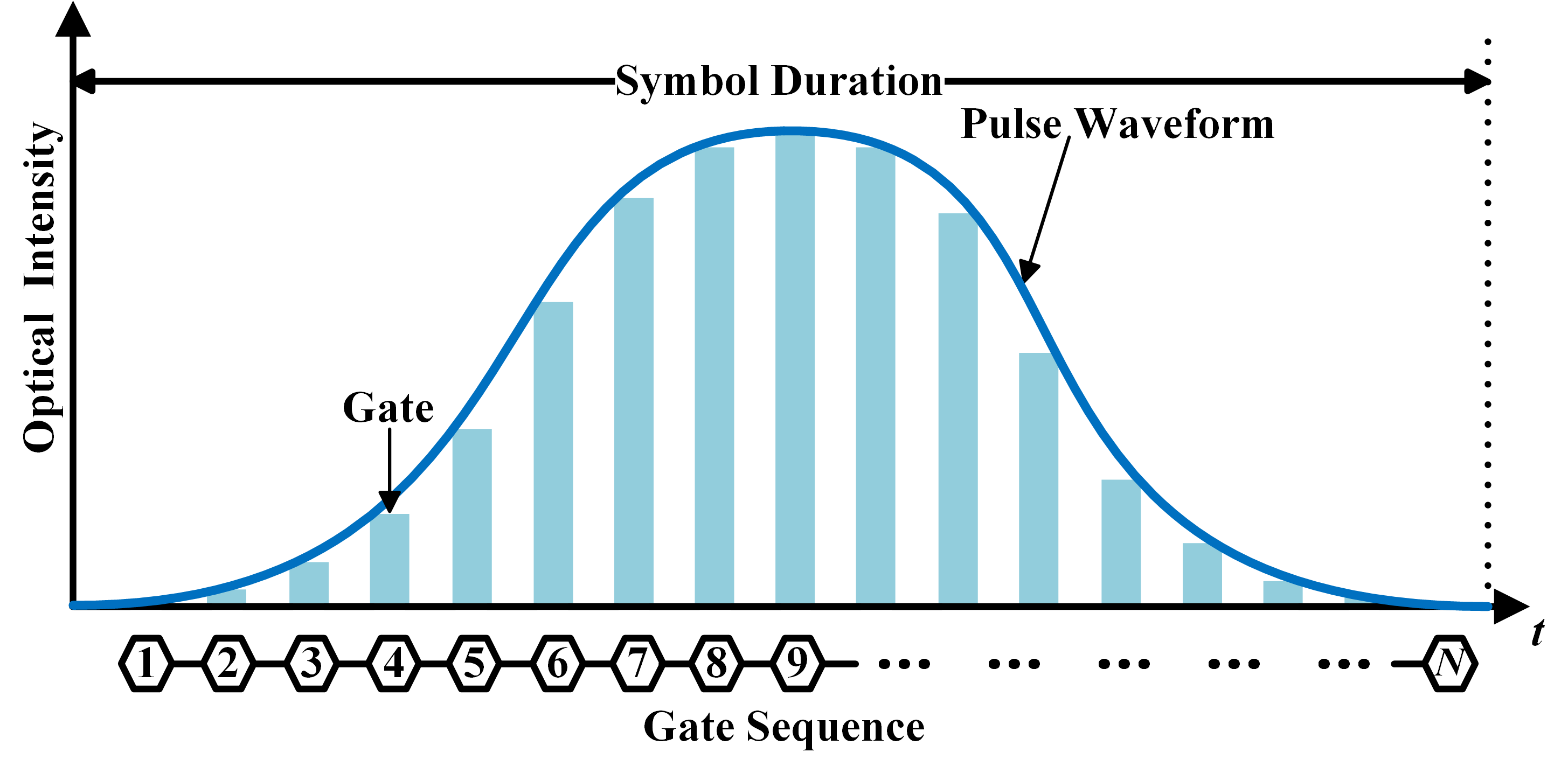}
\caption{The detection scheme for a photon-counting receiver designed for non-rectangle waveforms.}
\label{fig_1}
\end{figure}

Due to the weak light photons following a Poisson arrival process, the probability of $k$ photoelectrons being detected during a symbol duration $\left( {0,t} \right)$ is given by \cite{ref15}
\begin{equation}\label{eq1}
p\left( {k,t} \right) = \frac{{\left( {{\lambda _{\rm{p}}}t} \right){{\mathop{\rm e}\nolimits} ^{ - {\lambda _{\rm{p}}}{t}}}}}{{k!}}
\end{equation}
where the constant ${\lambda _{\rm{p}}}$ represents the average photoelectron arrival rate. Define ${\lambda _{\rm{p}}} \buildrel \Delta \over = {p _{\rm{de}}} \left( {{\lambda _{\rm{s}}} + {\lambda _{\rm{b}}}} \right) + {\lambda _{\rm{d}}}$, where ${p _{\rm{de}}} ,{\lambda _{\rm{s}}},{\lambda _{\rm{b}}},{\lambda _{\rm{d}}}$ denote the PDE, signal photon rate, background photon rate, and dark carrier rate, respectively.

Based on (\ref{eq1}), the trigger probability is given by \cite{ref6}

\begin{equation}\label{eq2}
P = 1 - p\left( {0,{\tau _{\rm{g}}}} \right) = 1 - {{\mathop{\rm e}\nolimits} ^{ - {\lambda _{\rm{p}}}{\tau _{\rm{g}}}}}
\end{equation}

For the $n$-th gate, the trigger probability is contingent on the photoelectron arrival rate and can be expressed as 
\begin{equation}\label{eq3}
{P_n} = 1 - \exp \left( { - \int_{t_n^{\rm{s}}}^{t_n^{\rm{e}}} {{\lambda _t}} {\rm{d}}t} \right)
\end{equation}
where $\left( {t_n^{\rm{s}},t_n^{\rm{e}}} \right)$  represents the start and the end times of the $n$-th gate, and ${\lambda _t} $ denote the photoelectron arrival rate at specific time instant $t$. Subsequently, by computing the trigger probabilities for all the gates within the symbol duration, denoted as $\left\{ {{P_1},{P_2}, \ldots ,{P_N}} \right\}$, where $N$ is the total number of gates in the symbol duration.

The photon counts detected within the symbol duration are considered as a random variable and expressed as follows:
\begin{equation}\label{eq4}
\lambda  = \sum\limits_{n = 1}^N {{I_n}} 
\end{equation}
where ${I_n}$ representing the gate state of the $n$-th gate, assumes the value "1" with probability ${P_n}$ and "0" with probability $1 - {P_n}$. This results in a Poisson binomial distribution with success probabilities $\left\{ {{P_1},{P_2}, \ldots ,{P_N}} \right\}$ that are not identical.

\subsection{DFT-CF Method}

\begin{figure*}[!t]
\centering
\subfloat[]{\includegraphics[width=3.0in]{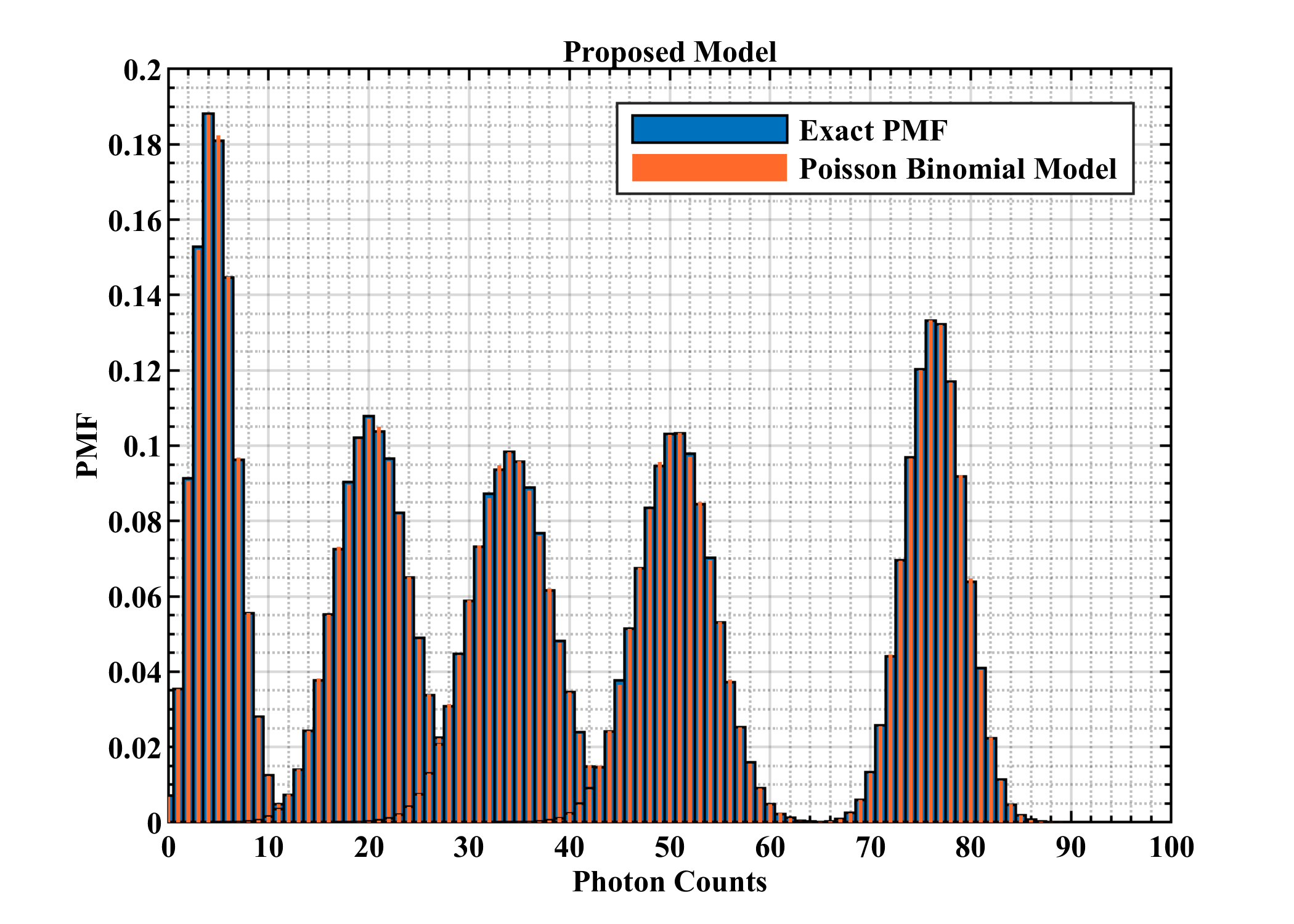}%
\label{fig_first_case}}
\hfil
\subfloat[]{\includegraphics[width=3.0in]{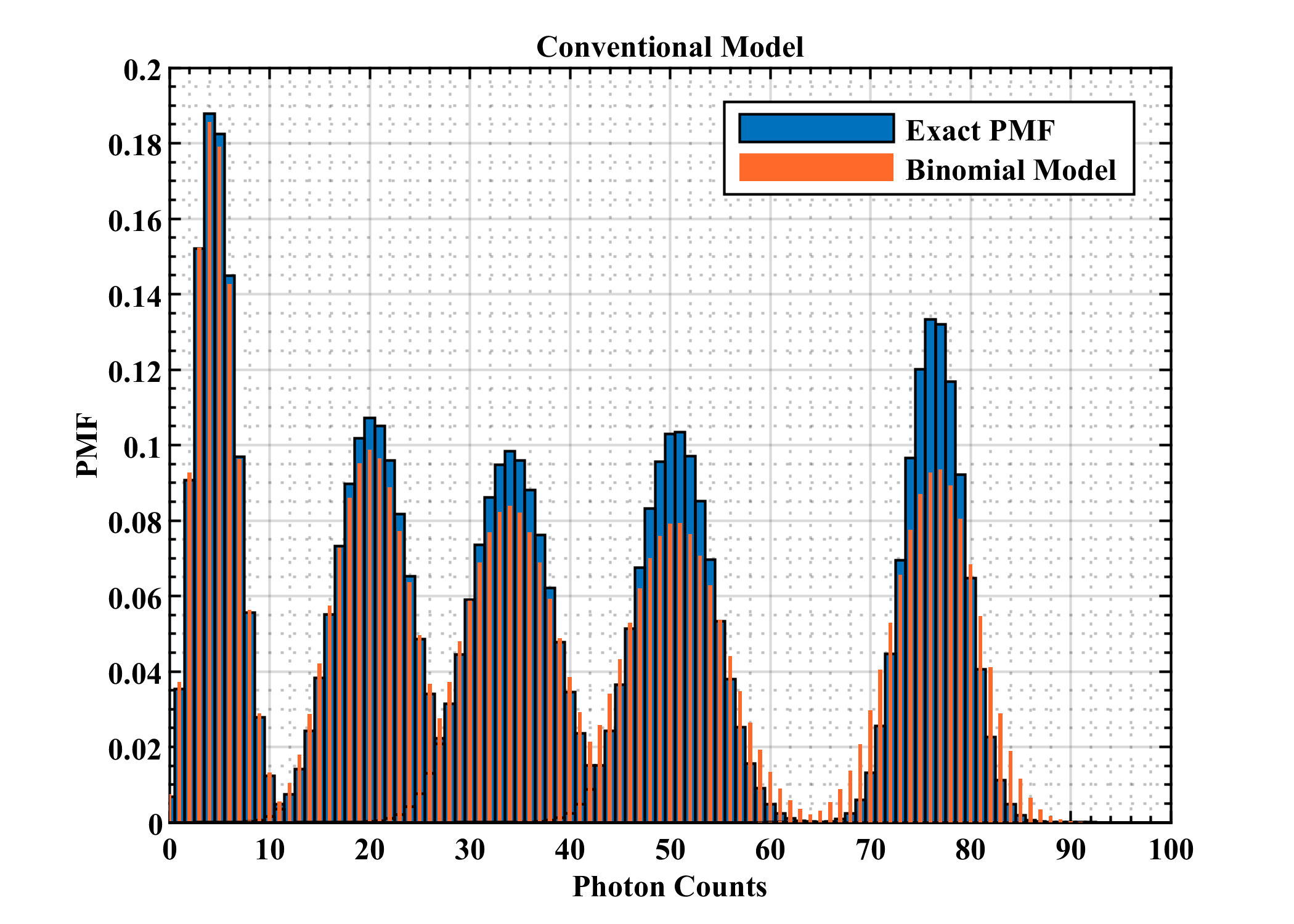}%
\label{fig_second_case}}
\caption{The analytical and exact PMFs of the detected photon counts for the Poisson binomial model and the conventional binomial model (${\lambda _{\text{s}}} = 0.1,0.5,1,2,8{{\text{c}} \mathord{\left/
 {\vphantom {{\text{c}} {{\text{ns}}}}} \right.
 \kern-\nulldelimiterspace} {{\text{ns}}}},N = 100$).}
\label{fig_sim}
\end{figure*}

To reduce the computational burden, we introduce the discrete Fourier transform characteristic function (DFT-CF) method to derive the probability mass function (PMF) of the detected photon counts. The characteristic function of a Poisson binomial random variable is derived as \cite{ref16}
\begin{align}\label{eq5}
\begin{gathered}
  \mathbb{E}\left[ {\exp \left( {it\lambda } \right)} \right] = \mathbb{E}\left[ {\exp \left( {it\sum\limits_{n = 1}^N {{I_n}} } \right)} \right] = \prod\limits_{j = 1}^N {\mathbb{E}\left[ {\exp \left( {it{I_j}} \right)} \right]}  \hfill \\
   = \prod\limits_{j = 1}^N {\left[ {1 - {P_j} + {P_j}\exp \left( {it} \right)} \right]}  = \sum\limits_{n = 0}^N {\Pr \left( {\lambda  = k} \right)\exp \left( {itn} \right)}  \hfill \\ 
\end{gathered} 
\end{align}
where  $i=\sqrt{-1} $. By applying the DFT to both sides of (\ref{eq5}), the complex-form PMF is obtained as

\begin{equation}\label{eq6}
\Pr \left( {\lambda  = k} \right) = \frac{1}{{N + 1}}\sum\limits_{l = 0}^N {\exp \left( { - i\omega lk} \right){x_l}} 
\end{equation}
where $\omega  = \tfrac{{2\pi }}{{N + 1}}$, ${x_l} = \prod\nolimits_{j = 1}^N {\left[ {1 - {P_j} + {P_j}\exp \left( {i\omega l} \right)} \right]}$. Using (\ref{eq6}), the PMF of the detected photon counts can be obtained based on the trigger probability sequence $\left\{ {{P_1},{P_2}, \ldots ,{P_N}} \right\}$.
To simplify calculations, $
{x_l}$ is expanded into its real parts and imaginary components as follows:
\begin{equation}\label{eq7}
\begin{gathered}
  {x_l} = \exp \left\{ {\sum\limits_{j = 1}^N {\ln \left[ {\left| {{z_j}\left( l \right)} \right|} \right]} } \right\}\left\langle {\cos \left\{ {\sum\limits_{j = 1}^N {{\text{Arg}}\left[ {{z_j}\left( l \right)} \right]} } \right\}} \right. \hfill \\
   + \left. {i\sin \left\{ {\sum\limits_{j = 1}^N {{\text{Arg}}\left[ {{z_j}\left( l \right)} \right]} } \right\}} \right\rangle  \hfill \\ 
\end{gathered} 
\end{equation}
where ${z_j}\left( l \right) = 1 - {P_j} + {P_j}\exp \left( {i\omega l} \right)$, and ${\text{arg}}\left[  \cdot  \right]$ represents the argument of a complex number. The complex computation of ${x_l}$ is then translated into real arithmetic as follows:
\begin{equation}\label{eq8}
\left| {{z_j}\left( l \right)} \right| = \sqrt {{{\left[ {1 - {P_j} + {P_j}\cos \left( {\omega l} \right)} \right]}^2} + {{\left[ {{P_j}\sin \left( {\omega l} \right)} \right]}^2}} 
\end{equation}

\begin{align}\label{eq9}
{\rm{arg}}\left[ {{z_j}\left( l \right)} \right] = \left\{ {\begin{array}{*{20}{c}}
{\left\{ {\begin{array}{*{20}{c}}
{\arctan \left( {\frac{{{P_j}\sin \left( {\omega l} \right)}}{{1 - {P_j} + {P_j}\cos \left( {\omega l} \right)}}} \right)}\\
{1 - {P_j} + {P_j}\cos \left( {\omega l} \right) > 0}
\end{array}} \right.}\\
{\left\{ {\begin{array}{*{20}{c}}
{{\rm{\pi }} + \arctan \left( {\frac{{{P_j}\sin \left( {\omega l} \right)}}{{1 - {P_j} + {P_j}\cos \left( {\omega l} \right)}}} \right)}\\
{{P_j}\sin \left( {\omega l} \right) \ge 0,1 - {P_j} + {P_j}\cos \left( {\omega l} \right) < 0}
\end{array}} \right.}\\
{\left\{ {\begin{array}{*{20}{c}}
{ - {\rm{\pi }} + \arctan \left( {\frac{{{P_j}\sin \left( {\omega l} \right)}}{{1 - {P_j} + {P_j}\cos \left( {\omega l} \right)}}} \right)}\\
{{P_j}\sin \left( {\omega l} \right) < 0,1 - {P_j} + {P_j}\cos \left( {\omega l} \right) < 0}
\end{array}} \right.}\\
{\left\{ {\begin{array}{*{20}{c}}
{\frac{{\rm{\pi }}}{2}}\\
{{P_j}\sin \left( {\omega l} \right) > 0,1 - {P_j} + {P_j}\cos \left( {\omega l} \right) = 0}
\end{array}} \right.}\\
{\left\{ {\begin{array}{*{20}{c}}
{ - \frac{{\rm{\pi }}}{2}}\\
{{P_j}\sin \left( {\omega l} \right) < 0,1 - {P_j} + {P_j}\cos \left( {\omega l} \right) = 0}
\end{array}} \right.}
\end{array}} \right.
\end{align}

According to (\ref{eq6}), the PMF of the detected photon counts is the inverse discrete Fourier transform (IDFT) of ${x_l}$. In practical computation, by applying the fast Fourier transform (FFT) to the sequence $\left\{ {{{{x_0}} \mathord{\left/
 {\vphantom {{{x_0}} {N + 1}}} \right.
 \kern-\nulldelimiterspace} {N + 1}},{{{x_1}} \mathord{\left/
 {\vphantom {{{x_1}} {N + 1}}} \right.
 \kern-\nulldelimiterspace} {N + 1}}, \ldots ,{{{x_N}} \mathord{\left/
 {\vphantom {{{x_N}} {N + 1}}} \right.
 \kern-\nulldelimiterspace} {N + 1}}} \right\}$, we obtain the PMF as the output of the FFT $\left\{ {\Pr \left( {\lambda  = 0} \right),\Pr \left( {\lambda  = 1} \right), \ldots ,\Pr \left( {\lambda  = N} \right)} \right\}$.
\subsection{PMF Comparisons}
To evaluate the analytical model, we use a Gaussian pulse waveform defined as follows:
\begin{equation}\label{eq10}
{f_m}\left( t \right) = {\lambda _{\rm{s}}}(m)\frac{6}{{\sqrt {2\pi } }}\exp \left\{ {\frac{{ - 18{{\left( {t - {{{T_{\rm{s}}}} \mathord{\left/
 {\vphantom {{{T_{\rm{s}}}} 2}} \right.
 \kern-\nulldelimiterspace} 2}} \right)}^2}}}{{{T_{\rm{s}}}^2}}} \right\}
\end{equation}
where ${f_m}\left( t \right)$ is the waveform function of "$m$" symbol, ${\lambda _{\rm{s}}}(i)$ is the average signal photon rate for "$m$" symbol, and $t$ is the time instant within the symbol duration $0 \le t \le {T_{\rm{s}}}$.

Fig. 2 illustrates the discrepancy between the exact and analytical PMFs of detected photon counts for a Gaussian pulse. In this figure, we consider a range of incident photon rates of ${\lambda _{\rm{s}}} = 0.1,0.5,1,2,8$ and a background radiation of ${\lambda _{\rm{b}}} = 0$. Fig. 2a shows that the PMFs obtained by the Poisson binomial model closely match the exact PMFs. In contrast, Fig. 2b shows a significant deviation between the binomial model and the exact results at medium and high photon counts. This indicates that the analytical Poisson binomial model accurately predicts the PMFs of detected photon counts for practical photon-counting receivers.

\section{Proposed Signal Estimation and Decision Method}
In photon-counting receivers, the optical signal intensity is estimated by counting photons detected within the symbol duration. We propose a method that involves retransmitting pilot symbols, where symbols "0", "1",..., "$M-1$" are retransmitted respectively. The estimation of the trigger probability for the $n$-th gate within "$m$" symbol duration is given by
\begin{equation}\label{eq11}
\hat P_n^m = \frac{{X_n^m}}{K}
\end{equation}
where $X_n^m$ and $K$ represents the number of detected photon counts and total gates for the $n$-th gate across all occurrences of "$m$" symbol. The superscript "$m$" denotes the symbol information. By calculating (\ref{eq11}), we can obtain the estimated trigger probability sequence $\left\{ {\hat P_1^m,\hat P_2^m, \ldots ,\hat P_N^m} \right\}$ for "$m$" symbol. 

\begin{algorithm}[H]
\caption{Signal Estimation and Decision}\label{alg:alg1}
\begin{algorithmic}
\STATE 
\STATE {\textbf{Input:} Detected photon counts $X$ and $X_n^m,N,{p _{\rm{de}}} ,{\lambda _{\text{s}}},{\lambda _{\text{b}}}$} 
\STATE {\textbf{Output:} Demodulated symbol information $Symbol$} 
\STATE \textbf{1:} Compute $\hat P_n^m$ using (\ref{eq11})
\STATE \textbf{2:} Substitute $\left\{ {\hat P_1^m,\hat P_2^m, \ldots ,\hat P_N^m} \right\}$ into (\ref{eq8}) (\ref{eq9}) to obtain $x_l^m$ 
\STATE \textbf{3:} Apply FFT to  $\left\{ {{{x_0^m} \mathord{\left/
 {\vphantom {{x_0^m} {N + 1}}} \right.
 \kern-\nulldelimiterspace} {N + 1}},{{x_1^m} \mathord{\left/
 {\vphantom {{x_1^m} {N + 1}}} \right.
 \kern-\nulldelimiterspace} {N + 1}}, \ldots ,{{x_N^m} \mathord{\left/
 {\vphantom {{x_N^m} {N + 1}}} \right.
 \kern-\nulldelimiterspace} {N + 1}}} \right\}$\STATE \textbf{ }\hspace{0.3cm} to obtain $\left\{ {{{\Pr }^m}\left( {\lambda  = 0} \right),{{\Pr }^m}\left( {\lambda  = 1} \right), \ldots ,{{\Pr }^m}\left( {\lambda  = N} \right)} \right\}$ 
\STATE \textbf{4: If} $X < \hat kt{h_0}$ \textbf{then} ${Symbol}={0}$
\STATE \textbf{5: }\hspace{0.3cm} \textbf{Else if} $
\hat kth\left( {{x_{m - 1}}} \right) \le X < \hat kth\left( {{x_m}} \right)$  \textbf{then} 
\STATE \hspace{2.3cm}  ${Symbol} = {m-1}$
\STATE \textbf{6:} \hspace{0.8cm} \textbf{Else if} $X \ge \hat kt{h_{M - 1}}$  \textbf{then} ${Symbol}={M-1}$
\STATE \textbf{7: End if}
\STATE \textbf{8: return} $Symbol$ 
\end{algorithmic}
\label{alg1}
\end{algorithm}

Using (\ref{eq6}), we set ${ {{\Pr }^m \left( {\lambda=kt{h_m}} \right)} }={{{\Pr }^{m-1} \left( {\lambda=kt{h_m}} \right)}}$, the thresholds can be obtained. However, this equation does not yield an analytical solution. Due to the unimodal characteristic of Poisson binomial distribution, we propose a straightforward method to determine the threshold using the maximum likelihood (ML) criterion. The proposed signal estimation and decision algorithm is presented in Algorithm \ref{alg:alg1}.

\section{Results and Discussions}
In this section, we compare the symbol error performance of the proposed algorithm with that of the conventional algorithm. The background photon rate is set to ${\lambda _{\text{b}}} = 0.01\sim2{{\text{c}} \mathord{\left/ {\vphantom {{\text{c}} {{\text{ns}}}}} \right.\kern-\nulldelimiterspace} {{\text{ns}}}}$, as measured in a previous experiment \cite{ref17}. Additionally, the square-root signaling technique is introduced to design a pulse amplitude modulation (PAM) signal constellation, which is suitable for signal-dependent noise systems \cite{ref18}. Building on square-root signaling, we modify the PAM signal constellation to $\left\{ {0,0.25{\lambda  _{\text{s}}},0.56{\lambda _{\text{s}}},{\lambda _{\text{s}}}} \right\}$ for 4-PAM. The expression for the optical pulse waveform is provided by (\ref{eq10}). The other simulation parameters are listed in Table \ref{tab:table1}.

\begin{table}[!t]
\caption{The setting of simulation parameters\label{tab:table1}}
\centering
\begin{tabular}{|c|c|c|}
\hline
Notation & Description & Value\\

\hline
$\eta $ & Photon detection efficiency & $10\% $\\

\hline 
${\tau _{\text{g}}}$ & Gate-ON interval& $2{\text{ns}}$\\

\hline 
${\tau _{{\text{cyc}}}}$ & Dead time & $8{\text{ns}}$\\

\hline 
${\lambda _{\text{s}}}$ & Signal photon rate & $1\sim15{{\text{c}} \mathord{\left/
 {\vphantom {{\text{c}} {{\text{ns}}}}} \right.
 \kern-\nulldelimiterspace} {{\text{ns}}}}$\\

\hline 
${\lambda _{\text{b}}}$ & Background photon rate & $0.01\sim2{{\text{c}} \mathord{\left/
 {\vphantom {{\text{c}} {{\text{ns}}}}} \right.
 \kern-\nulldelimiterspace} {{\text{ns}}}}$\\
 
\hline 
${\lambda _{\text{d}}}$ & Dark carrier rate & $4.4 \times {10^{ - 5}}{{\text{c}} \mathord{\left/
 {\vphantom {{\text{c}} {{\text{ns}}}}} \right.
 \kern-\nulldelimiterspace} {{\text{ns}}}}$\\
 
\hline
\end{tabular}
\end{table}

\begin{figure}[!t]
\centering
\includegraphics[width=3.0in]{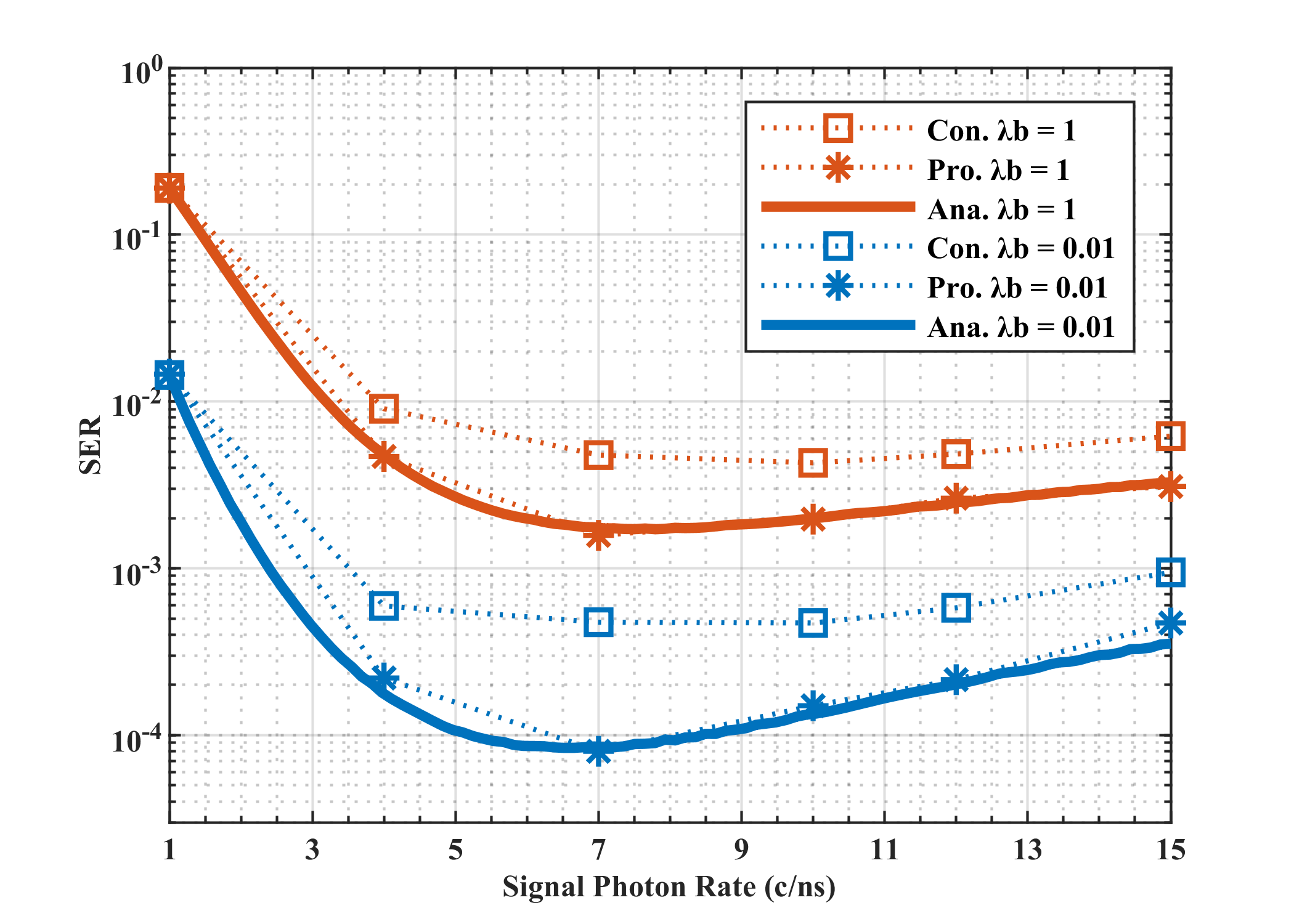}
\caption{Comparison of the SERs for the proposed algorithm with that of the conventional algorithm, as a function of optical signal intensity ($N = 400$).}
\label{fig_3}
\end{figure}
Fig. 3 shows that the SER of the proposed algorithm surpasses that of the conventional algorithm, which relies on the binomial model. In the medium and high optical regimes (${\lambda _{\text{s}}}>3$), the SER of the Poisson binomial model outperforms that of the binomial model by one order of magnitude. Remarkably, when the receiving optical signal intensity nears its optimal level, the improvement in SER exceeds one order of magnitude. This demonstrates that the proposed algorithm significantly enhances error performance.

\begin{figure}[!t]
\centering
\includegraphics[width=3.0in]{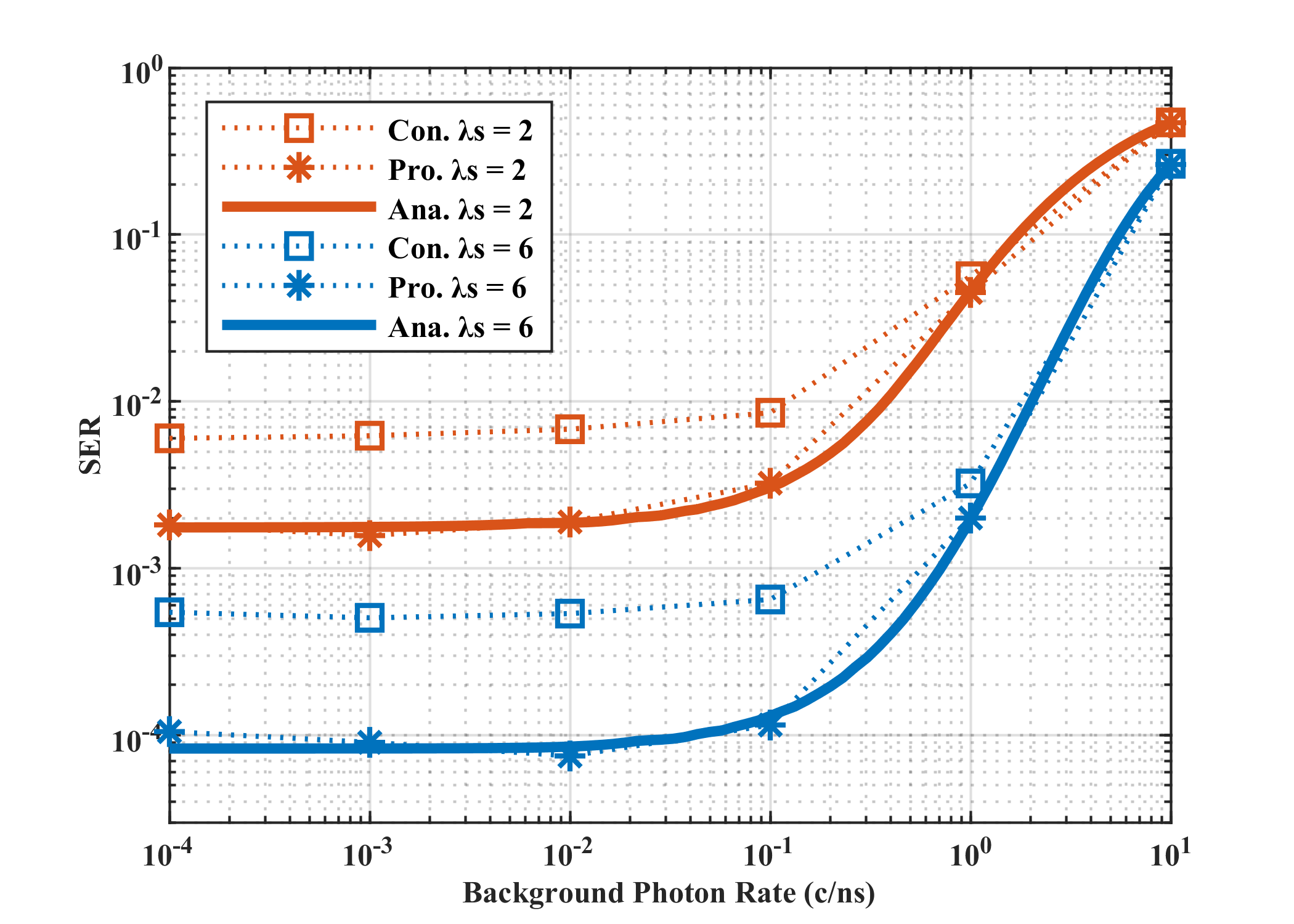}
\caption{Comparison of the SERs for the proposed algorithm with that of the conventional algorithm, as a function of background radiation ($N = 400$).}
\label{fig_4}
\end{figure}

As depicted in Fig. 4, the SER improvement accelerates rapidly as background radiation decreases. Specifically, when the background radiation falls below the tolerated level (${\lambda _{\text{b}}} < {10^{ - 1}}$), the SER is reduced by one order of magnitude. Together with Fig. 3, these results clearly show that the proposed algorithm significantly improves SER both at optimal receiving intensity. This improvement becomes less discernible in extreme optical regimes where the trigger probabilities approach 0 or 1, making the differences between gates negligible. Consequently, the SER improvement achieved by Poisson binomial model becomes indiscernible. In practical photon-counting receivers, it is crucial to precisely control the incident signal photon rate at its optimal level.

\section{Conclusion}
In conclusion, we have conducted a comprehensive study on signal detection method for practical photon-counting receivers. Based on this, we proposed a novel signal estimation and decision algorithm that outperforms conventional algorithm. Overall, this study contributes to the advancement of signal processing techniques for photon-counting receivers.

\vfill

\end{document}